\documentclass{emulateapj}

\usepackage{natbib}
\begin{document}

\title{The 3$\mu$m spectrum of Jupiter's irregular satellite Himalia}
\author{M.E. Brown}
\affil{Division of Geological and Planetary Sciences, California Institute
of Technology, Pasadena, CA 91125}
\email{mbrown@caltech.edu}
\author{A.R. Rhoden}
\affil{Johns Hopkins University Applied Physics Laboratory, Laurel, MD 20723}
\email{Alyssa.Rhoden@jhuapl.edu}

\begin{abstract}
We present a medium resolution spectrum of Jupiter's irregular satellite
Himalia covering the critical 3 $\mu$m spectral region. The spectrum
shows no evidence for aqueously altered phyllosilicates, as had been
suggested from the tentative detection of a 0.7 $\mu$m absorption, but
instead shows a spectrum strikingly similar to the C/CF type asteroid
52 Europa. 52 Europa is the prototype of a class of asteroids generally
situated in the outer asteroid belt between less distant asteroids which
show evidence for aqueous alteration and more distant asteroids which
show evidence for water ice. The spectral match between Himalia and 
this group of asteroids is surprising and difficult to reconcile with
models of the origin of the irregular satellites.
\end{abstract}

\section{Introduction}
The origin of the irregular satellites of the giant planets
-- small satellites in distant, 
eccentric, and inclined orbits about their parent body -- remains
unclear. Early work suggested that the objects were captured from
previously heliocentric orbits by gas drag \citep{1979Icar...37..587P}
or collisions \citep{1971Icar...15..186C},
while more recent ideas have suggested that the irregular satellites
were captured from the outer solar system during the final
stages of a giant planet dynamical instability which violently rearranged
the solar system \citep{2007AJ....133.1962N,2014ApJ...784...22N}. 

The surface compositions of the irregular satellites have often been
used to attempt to understand the origins of these bodies. Unfortunately,
the information we have on their compositions is sparse. Visible through near-infrared
photometry and spectroscopy show generally featureless spectra with 
flat to slightly red slopes in the optical and slightly red slopes in
the near-infrared 
\citep{2001Icar..154..313R,2003Icar..166...33G,2004ApJ...605L.141G, 2006Icar..180..453V}. 
By analogy to asteroids, these characteristics 
often lead to these satellites to be referred to as C-type or P/D-type objects, 
with the implication
of a common origin between the irregular satellites and these 
asteroids, though the genetic
connection is far from
clear.

Himalia is the largest of the Jovian irregular satellites and the
largest member of the prograde group of Jovian irregular satellites,
which are all thought to be part of a single collisional family. 
The spectra of Himalia and most of the other
members of its family show a distinct broad absorption feature
from the visible to the near infrared with a center around 1 $\mu$m 
\citep{2003Icar..166...33G,2004ApJ...605L.141G}.
Detection of a 0.7 $\mu$m absorption band has been reported from some data,
but not from others \citep{2000Icar..145..445J,2006Icar..180..453V}. 
Such a feature, if actually present,
may result from
oxidized iron in phyllosilicate minerals 
potentially caused by aqueous processing
on these bodies. 

In a study of dark asteroids in the outer belt \citet{2012Icar..219..641T} 
found that all
asteroids that they observed which have clear 0.7 $\mu$m absorption bands
showed a sharp 3 $\mu$m absorption feature consistent with hydroxyl-group 
absorption in phyllosilicates.
A spectrum of Himalia obtained from a distant encounter
of the Cassini spacecraft shows the possibility of an absorption 
at 3 $\mu$m, but the low signal-to-noise and low spectral resolution prohibits 
clear interpretation \citep{2004Icar..172..163C}. 

Here we present a high resolution moderate signal-to-noise spectrum of Himalia
from 2-4 $\mu$m obtained from the Keck Observatory. The spectrum shows
unambiguous evidence of an absorption at 3 $\mu$m. Below we identify and
model the absorption feature and discuss the 
possible surface composition of Himalia. Finally we discuss the implication
for the origin of the parent body of Himalia and for the irregular satellites.

\section{Observations}
We observed Himalia on the nights of 26 and 28 November 2013 using the
facility moderate resolution near infrared spectrograph NIRSPEC
\citep{1998SPIE.3354..566M}. NIRSPEC
in its current configuration covers a limited spectral range, so to cover
the full 2.2-3.8 $\mu$m region requires two spectral settings, one covering 
2.25 - 3.10 $\mu$m and the other covering 2.96 - 3.81 $\mu$m. In each spectral
setting, we nodded the target between two positions on the 0.57 arcsecond 
wide slit. For the shorter wavelength setting integration times were 7.5 
seconds with 30 coadds, and we obtained 12 such spectra each night for
a total integration time of 45 minutes. For the
the longer wavelengths, with higher thermal
emission, integration times were 1 second with 150 coadds, and we obtained
24 such spectra the first night and 36 the second, for a total integration
time of 150 minutes. We obtained identical spectral of G 91-3, a G2V
star with $V=7.4$ a distance a 3.2 degrees from Himalia. All Himalia 
spectra were obtained at airmasses between 1.0 and 1.1, and the spectra
of the calibrator star were obtained within 0.1 of the same airmass, and 
immediately preceding or following the Himalia observation.

To obtain calibrated data from the raw spectra, we first subtract
each temporally adjacent pair of images in which the spectra are
in two separate spatial positions. This subtraction removes the bulk
of the telescope and sky emission, leaving a positive and negative
spectrum of the source along with any small sky or telescope changes
that occurred between the two spectra.
A point-source spectrum projects as a curve onto the detector, and
sky emission lines which fill the slit have variable curvature
as a function of wavelength. 
We use measurements
of the positions of the bright calibrator star in the two slit positions
and measurements of the positions of sky lines in the longer target
exposures to create a template allowing us to map the non-rectilinear 
geometry of the data onto a new rectilinear grid with the spatial dimension
along one axis and the wavelength along the other. We then use this 
template to remap the spectral pairs onto a rectilinear grid. We find
the spatial profile of the spectrum by taking a median along the spectral
dimension. Typical full-width at half-maximum (FWHM) in the spatial dimension
is 0.9 arcseconds. To obtain the flux at each wavelength we 
fit the expected spatial profile plus a linear offset to 50 pixels
on either side of the maximum of the spatial profile. Bad pixels
are flagged and ignored in the fit. This procedure
removes any residual sky lines and also corrects for most
of the thermal
drift of the telescope which can be important beyond about 3.5 $\mu$m. 
Examining the final spectral images individually, we found, however, that
for about 20\% of the longer wavelength spectra the pair 
subtracted poorly, leaving large residuals in the spatial dimension,
which are most likely caused by thermal changes in the telescope. We
discard those data affected by this problem, leaving a final total integration
time for the longer wavelength data of 122.5 minutes. The individual
spectra are summed with 5$\sigma$ and greater outliers from the mean
at each wavelength removed. An identical procedure is performed for
the stellar calibration spectra, and the target spectrum is divided by
the stellar spectrum to obtain a relatively calibrated spectrum. 
The shorter and longer wavelength spectra are scaled such that the median
of the data in the overlap region is identical. Wavelength calibration
is obtained by matching the spectrum of the calibrator to a theoretical
transmission spectrum of the terrestrial atmosphere. 

The original data has a spectral resolution of approximately 
R/$\Delta$R$\sim$2000. To increase the signal-to-noise per resolution element, 
we convolve the spectrum with a gaussian filter with a FWHM of 8 pixels,
for a final 2-pixel resolution of $\sim$500. To achieve the strongest
possible constraint on any 3 $\mu$m absorption, we separately 
calculate reflectances in windows centered at 2.828 and
2.857 $\mu$m, which present small transparent regions through the
atmosphere surrounded by strong telluric absorption.
 The final calibrated spectrum
is shown in Figure 1. While the spectrum is only calibrated in terms of
relative reflectance, we estimate the reflectance by assuming a
value of for the K-band albedo of 0.08 after \citet{2004Icar..172..163C}.

\section{Results}
The spectrum of Himalia shows a clear sign of absorption in the 3$\mu$m
region (Fig. 1). Because of the report of the 0.7 $\mu$m absorption feature and
its associated with hydrated silicates, we first attempt
to model the 3 $\mu$m spectrum of Himalia with typical hydrated silicate
spectra. These spectra have peak absorption in the unobserved 2.5 to 2.8 $\mu$m 
region and their reflectivities sharply rise longward of 2.8 $\mu$m 
(see the \citet{2012Icar..219..641T} "sharp" spectral group). In Figure 1 we 
compare Himalia to a spectrum of a CM chondrite measured under dry
conditions \citep{2013M&PS...48.1618T} whose spectrum is
consistent with
cronstedtite 
-- a phyllosilicate which is a major constituent of CM chondrites -- scaled
to match the continuum level at wavelengths shortward of 2.5 $\mu$m.
While the match beyond 3.0 $\mu$m is adequate,
shortward of 3.0 $\mu$m the spectrum of Himalia turns upward rather
than continuing downward like a typical phyllosilicate. The spectrum of
Himalia fits neither this particular mineral nor the general characteristics
of the \citet{2012Icar..219..641T} "sharp" class of dark asteroids with 0.7 $\mu$m
absorptions and sharply rising spectra from 2.8 $\mu$m.

The absorption centered at approximately 3.0 $\mu$m in the spectrum of Himalia
appears more similar to water ice than to OH absorption. To examine this
possibility we construct a Shkuratov model \citep{1999Icar..137..235S}
of dark grains covered
with fine grained ice, following the prescription of 
\citet{2010Natur.464.1322R}. 
In our model the dark grains have optical constants chosen
to match the overall albedo level of the continuum
and the ice is a fine-grained frost
with a volume fraction of 1.5\% (chosen to match the depth of the absorption)
coating the dark grains. Water ice optical
constants are obtained from \citet{2009ApJ...701.1347M}.
This model provides an adequate match to the
2.8 to 3.1 $\mu$m range of the data, but Himalia shows significantly more
absorption out to 3.6 $\mu$m which cannot be accounted for by
water ice (Figure 1).
\begin{figure}
\plotone{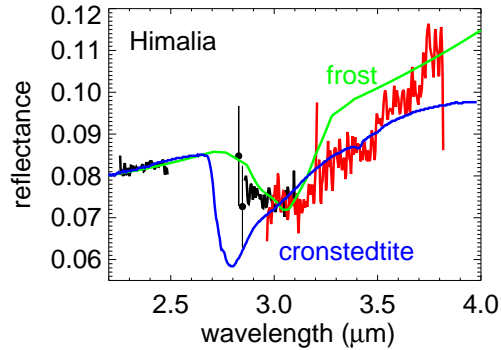}
\caption{The spectrum of Himalia from 2.2 to 3.8 $\mu$m. The shorter and
longer wavelength NIRSPEC settings are shown as black and red, respectively.
The spectrum is compared to a CM chondrite with a spectrum consistent with
cronstedtite (blue) -- a commonly occurring meteoritic phyllosilicate -- and to
a model of fine grained water ice frost overlying dark grains (green). Neither
material provides a satisfactory match to the spectrum of Himalia.
}
\end{figure}

A search of all available libraries of reflectance spectra reveals no materials
which can match the spectrum of Himalia. A closer match can be obtained, however,
by comparing the spectrum of Himalia with those of dark main belt asteroids.
The spectrum of asteroid 24 Themis, for example, similarly contains a water ice-like
feature at $\sim$3.1 $\mu$m followed by continued absorption out to 3.6 $\mu$m
\citep{2010Natur.464.1320C,2010Natur.464.1322R}, but a close comparison of
the two spectra shows that the peak of absorption in the
spectrum of Himalia is shifted by $\sim$0.05 $\mu$m compared to that of
24 Themis. 
An excellent match to the spectrum of Himalia comes, however, 
from the spectrum of the CF type asteroid 52 Europa obtained by
\citet{2012Icar..219..641T}. In Figure 2 we 
compare the spectrum 
of Himalia to a model where we have taken the spectrum of 52 Europa and 
allowed the spectral slope and depth of absorption to be free parameters
in a fit to the data. The best fit gives a spectrum with a 2.4 to 3.7 $\mu$m
slope 23\% 
redder than that of 52 Europa and an absorption 70\% deeper.
\begin{figure}
\plotone{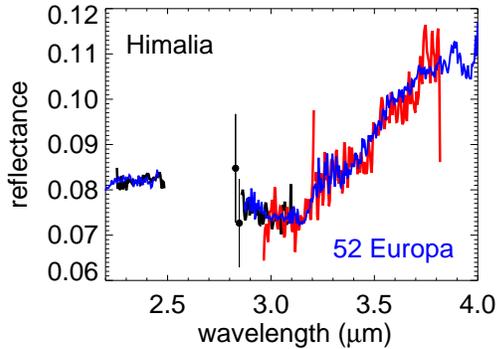}
\caption{
The spectrum of Himalia compared to the spectrum of the 52 Europa (blue), the 
prototype of the \citet{2012Icar..219..641T} "Europa-like" spectrum.
The match is excellent, as is the match to the shorter wavelength
portions of the spectrum. The surface composition of the Europa-like
group is unknown, but the other members are all situated in the middle
of the asteroid belt.
}
\end{figure}
The resulting spectrum is nearly indistinguishable
from that of Himalia. Moreover, the overall shape of the visible through
and 2 $\mu$m spectrum with a broad absorption
centered at $\sim$1.2 $\mu$m appears very similar to the available 
broadband photometry of 
Himalia \citep{2003Icar..166...33G,2004ApJ...605L.141G}. 
However, the spectra of the Europa-like group of asteroids
lack any 0.7 $\mu$m absorption
feature. Based on the otherwise strong match between Himalia and 52 Europa and
the strong correlation between 0.7 $\mu$m absorption and the appearance of
the sharp OH-like absorption at 3 $\mu$m {\it not} seen on Himalia, we 
suspect that no actual 0.7 $\mu$m absorption on Himalia is present. This
suspicion requires observational confirmation.
 We conclude that had Himalia been a dark asteroid
in the survey of \citet{2012Icar..219..641T} 
it would have clearly fit into the 
group of Europa-type spectra that includes 52 Europa, 31 Euphrosyne, and 451 Patientia.

\section{Discussion}
The Jupiter irregular satellite Himalia has a 0.5 to 3.8 $\mu$m
spectrum which fits into one of the four spectral categories identified by 
\citet{2012Icar..219..641T} for dark asteroids in the outer asteroid belt.
Himalia appears to have a "52 Europa-like" spectrum. The 3 known asteroids 
with Europa-like spectra are all in the middle part of the asteroid belt,
with semimajor axes between 3.1 and 3.2 AU. The majority of the 
asteroids surveyed with semimajor axes inward of 3.1 AU had evidence
for OH bearing phyllosilicates suggestive of processing by liquid water, while
the majority of more distant asteroids had rounded 3 $\mu$m spectra which
appear to be evidence for the continued presence of water ice rather than
hydrated silicates. The surface composition of the asteroids
with Europa-like spectra is unclear, 
though their semimajor axes between the regions of the asteroid belt
containing the other two types 
of spectra suggests that they might be some sort of transition objects.

In this context, the similarity of the spectrum of Himalia with 
those of the Europa-like group is surprising. If Himalia is genetically
related to the dark asteroids it might be expected to be like the most distant
of those asteroids, which tend to have rounded, ice-like spectra. Alternatively,
if the Jovian irregular satellites are instead derived from the same source
as the Kuiper belt objects, the surface spectral similarity with objects in
the mid-section of the asteroid belt points to either compositional or
surface similarities between these populations or both.

Further insight into the origin of this spectral similarity will come from
the discovery of additional objects with
Europa-like spectra in the asteroid belt. With
only 3 objects currently known the true extent of their distribution is
unclear. Additional insight could also come from similar observations of Jupiter
trojan asteroids. In the context of the Nice model, these objects are also
thought to derive from the Kuiper belt source region, and they are currently
situated at the same semi-major axis and thus have the same thermal and
irradiation environment as the Jovian irregular satellites. 
An extensive visible to near-infrared survey of these objects found none
with the broad $\sim$1 $\mu$m absorption seen on Himalia, so it is clear that
these objects are not spectrally identical to Himalia. Nonetheless, examining 
their spectral morphology at 3 $\mu$m should provide important insight
into the genetic links and the surface processing of the varied populations
of the middle part of the solar system.

\acknowledgements
We thank Andy Rivkin and Josh Emery for illuminating conversations about the
nature of 3 $\mu$m absorptions in dark asteroids and for providing the spectra
of Themis and Europa. Driss Takir made excellent suggestions which 
improved the paper. This research was supported by Grant NNX09AB49G from
the NASA Planetary Astronomy Program.
A. Rhoden was partially supported through an appointment to the NASA Postdoctoral Program at the NASA Goddard Space Flight Center, administered by
Oak Ridge Associated Universities.
The data presented herein were obtained at the W.M. Keck Observatory, 
which is operated as a scientific partnership among the California Institute 
of Technology, the University of California and the National Aeronautics and 
Space Administration. The Observatory was made possible by the generous 
financial support of the W.M. Keck Foundation. The authors wish to 
recognize and acknowledge the very significant cultural role and reverence 
that the summit of Mauna Kea has always had within the indigenous 
Hawaiian community.  We are most fortunate to have the opportunity to 
conduct observations from this mountain.


\begin{thebibliography}{17}
\expandafter\ifx\csname natexlab\endcsname\relax\def\natexlab#1{#1}\fi

\bibitem[{{Campins} {et~al.}(2010){Campins}, {Hargrove}, {Pinilla-Alonso},
  {Howell}, {Kelley}, {Licandro}, {Moth{\'e}-Diniz}, {Fern{\'a}ndez}, \&
  {Ziffer}}]{2010Natur.464.1320C}
{Campins}, H., {Hargrove}, K., {Pinilla-Alonso}, N., {Howell}, E.~S., {Kelley},
  M.~S., {Licandro}, J., {Moth{\'e}-Diniz}, T., {Fern{\'a}ndez}, Y., \&
  {Ziffer}, J. 2010, \nat, 464, 1320

\bibitem[{{Chamberlain} \& {Brown}(2004)}]{2004Icar..172..163C}
{Chamberlain}, M.~A. \& {Brown}, R.~H. 2004, Icarus, 172, 163

\bibitem[{{Colombo} \& {Franklin}(1971)}]{1971Icar...15..186C}
{Colombo}, G. \& {Franklin}, F.~A. 1971, Icarus, 15, 186

\bibitem[{{Grav} \& {Holman}(2004)}]{2004ApJ...605L.141G}
{Grav}, T. \& {Holman}, M.~J. 2004, \apjl, 605, L141

\bibitem[{{Grav} {et~al.}(2003){Grav}, {Holman}, {Gladman}, \&
  {Aksnes}}]{2003Icar..166...33G}
{Grav}, T., {Holman}, M.~J., {Gladman}, B.~J., \& {Aksnes}, K. 2003, Icarus,
  166, 33

\bibitem[{{Jarvis} {et~al.}(2000){Jarvis}, {Vilas}, {Larson}, \&
  {Gaffey}}]{2000Icar..145..445J}
{Jarvis}, K.~S., {Vilas}, F., {Larson}, S.~M., \& {Gaffey}, M.~J. 2000,
  Icarus, 145, 445

\bibitem[{{Mastrapa} {et~al.}(2009){Mastrapa}, {Sandford}, {Roush},
  {Cruikshank}, \& {Dalle Ore}}]{2009ApJ...701.1347M}
{Mastrapa}, R.~M., {Sandford}, S.~A., {Roush}, T.~L., {Cruikshank}, D.~P., \&
  {Dalle Ore}, C.~M. 2009, \apj, 701, 1347

\bibitem[{{McLean} {et~al.}(1998){McLean}, {Becklin}, {Bendiksen}, {Brims},
  {Canfield}, {Figer}, {Graham}, {Hare}, {Lacayanga}, {Larkin}, {Larson},
  {Levenson}, {Magnone}, {Teplitz}, \& {Wong}}]{1998SPIE.3354..566M}
{McLean}, I.~S., {Becklin}, E.~E., {Bendiksen}, O., {Brims}, G., {Canfield},
  J., {Figer}, D.~F., {Graham}, J.~R., {Hare}, J., {Lacayanga}, F., {Larkin},
  J.~E., {Larson}, S.~B., {Levenson}, N., {Magnone}, N., {Teplitz}, H., \&
  {Wong}, W. 1998, in Society of Photo-Optical Instrumentation Engineers (SPIE)
  Conference Series, Vol. 3354, Infrared Astronomical Instrumentation, ed.
  A.~M. {Fowler}, 566--578

\bibitem[{{Nesvorn{\'y}} {et~al.}(2014){Nesvorn{\'y}}, {Vokrouhlick{\'y}}, \&
  {Deienno}}]{2014ApJ...784...22N}
{Nesvorn{\'y}}, D., {Vokrouhlick{\'y}}, D., \& {Deienno}, R. 2014, \apj, 784,
  22

\bibitem[{{Nesvorn{\'y}} {et~al.}(2007){Nesvorn{\'y}}, {Vokrouhlick{\'y}}, \&
  {Morbidelli}}]{2007AJ....133.1962N}
{Nesvorn{\'y}}, D., {Vokrouhlick{\'y}}, D., \& {Morbidelli}, A. 2007, \aj, 133,
  1962

\bibitem[{{Pollack} {et~al.}(1979){Pollack}, {Burns}, \&
  {Tauber}}]{1979Icar...37..587P}
{Pollack}, J.~B., {Burns}, J.~A., \& {Tauber}, M.~E. 1979, Icarus, 37, 587

\bibitem[{{Rettig} {et~al.}(2001){Rettig}, {Walsh}, \&
  {Consolmagno}}]{2001Icar..154..313R}
{Rettig}, T.~W., {Walsh}, K., \& {Consolmagno}, G. 2001, Icarus, 154, 313

\bibitem[{{Rivkin} \& {Emery}(2010)}]{2010Natur.464.1322R}
{Rivkin}, A.~S. \& {Emery}, J.~P. 2010, \nat, 464, 1322

\bibitem[{{Shkuratov} {et~al.}(1999){Shkuratov}, {Starukhina}, {Hoffmann}, \&
  {Arnold}}]{1999Icar..137..235S}
{Shkuratov}, Y., {Starukhina}, L., {Hoffmann}, H., \& {Arnold}, G. 1999,
  Icarus, 137, 235

\bibitem[{{Takir} \& {Emery}(2012)}]{2012Icar..219..641T}
{Takir}, D. \& {Emery}, J.~P. 2012, Icarus, 219, 641

\bibitem[{{Takir} {et~al.}(2013){Takir}, {Emery}, {McSween}, {Hibbitts},
  {Clark}, {Pearson}, \& {Wang}}]{2013M&PS...48.1618T}
{Takir}, D., {Emery}, J.~P., {McSween}, H.~Y., {Hibbitts}, C.~A., {Clark},
  R.~N., {Pearson}, N., \& {Wang}, A. 2013, Meteoritics and Planetary Science,
  48, 1618

\bibitem[{{Vilas} {et~al.}(2006){Vilas}, {Lederer}, {Gill}, {Jarvis}, \&
  {Thomas-Osip}}]{2006Icar..180..453V}
{Vilas}, F., {Lederer}, S.~M., {Gill}, S.~L., {Jarvis}, K.~S., \&
  {Thomas-Osip}, J.~E. 2006, Icarus, 180, 453

\end{thebibliography}
\end{document}